\documentclass{ws-p8-50x6-00}

\def \beq {\begin{equation}}
\def \eeq {\end{equation}}

\begin{document}
%
\title{Reply to a Comment of article \hspace{5cm}
``Evidence for Neutrinoless Double Beta Decay''}
\author{H.V. Klapdor-Kleingrothaus$^{1,2}$}

\address{$^{1}$Max-Planck-Institut f\"ur Kernphysik,
 	Postfach 10 39 80, D-69029 Heidelberg, Germany\\
	$^{2}$Spokesman of the HEIDELBERG-MOSCOW and GENIUS Collaborations,\\ 
	e-mail: klapdor@gustav.mpi-hd.mpg.de,\\
	 home page: http://www.mpi-hd.mpg.de/non$\_$acc/}

\maketitle              

	After our publication of the observed first indication 
	for neutrinoless double beta decay in 
\cite{KK-Evid01},
	a number of papers have been published 
\cite{KK-Sar_Evid01,KK2-Sar_Evid01,KK3-Sar_Evid01,Ma02-Ev,Barg02-Ev,Ueh02-Ev,Min-Evid02,Chat-Evid02,Xing-Evid02,Ham-Evid02,Ahl-Kirch-Evid02,Ma-Evid2-02,Hab-Suz02,Mohap02-Evid,Kays02-Evid,Fodor02-Evid,Pakv02-Evid,Rodej02-Evid,Kowal02,Niels02,He02-Evid,Josh02-Evid,Bran02-Evid,Cheng02-Evid,Smirn02-Evid,Wiss-Evid02}
	discussing its implications.
	In addition a 'Comment' has 
	been put to the Internet February 7, 2002,
\cite{Comm-Evid02} ({\it version 1})
	critisizing our publication 
\cite{KK-Evid01}. 
	(Meanwhile six of the nine claims in the 'Comment' 
	have been withdrawn 
\cite{Comm-Evid02} ({\it version 3}).)
	It is the purpose of this note to show that 
	none of their claims is justified (the same is true, 
	from the same arguments, for the criticism raised recently in 
\cite{Harn-Id02}).

	We shall show 
	this by answering this 'Comment' point by point.\\

	{\sf Point of Criticism 1.} 
	{\it 'There is no discussion of how a variation of the size 
	of the chosen analysis window would effect the significance 
	of the hypothetical peak'.} \\
	
	\underline{\sf Reply:}~~~ {\bf This is not true}. \\

	{\bf Figs. 4-6 of the Letter 
\cite{KK-Evid01}
	show the difference obtained for the probability of the signal 
	when choosing a large (2000-2080\,keV) window 
	and a small (2030-2048 and 2032-2046\,keV) window.
 
	In high-energy physics the usual procedure 
	in searches for resonances is to analyse an interval of  
	about $\pm$5$\sigma$ around the lines. 
	Our small window is in accordance with this.} 

	{\bf Moreover, computer-generated spectra 
	using Poisson random number generators for the background 
	and Gaussian random number generators for a line show that 
	a $\pm$(4-5)$\sigma$ analysis window around a line 
	allows for a reliable analysis (see 
\cite{DARK2002,DM2002}).  
	Details will be published in a forthcoming paper.}\\
	
	{\sf Point of Criticism 2.}
	{\it 'There is no relative peak strength analysis of 
	all the $^{214}{Bi}$ peaks. Quantitative yield evaluations 
	should be made on the low $^{214}{Bi}$ peaks 
	in the region of interest.'} \\

	They add in their section ``Relative Strength of the Bi peaks'' 
	and in their conclusion: 
	{\it `` a simple analysis of the $^{214}{Bi}$ peaks demonstrates 
	that the peak finding procedure used by KDHK produced spurious 
	peaks near the $\beta\beta$(0$\nu$) endpoint.''}\\

	\underline{\sf Reply:}~~~ \\

	{\bf The estimates of the $^{214}{Bi}$ intensities, 
	the authors of 
\cite{Comm-Evid02}
	present in their section 3, are 
	not correct for two reasons. 
	The first reason is due to the normalization the authors of
\cite{Comm-Evid02}
	derived from Fig.1 in 
\cite{HDM01},
	which unfortunately contains a normalisation error 
	of about a factor of 9 (see point 5). 
	The other reason is that they did not include 
	summing effects. 
	To understand relative intensities knowledge about 
	the location of the impurities inside the experimental setup
	is required. 
	This can be taken into account only by simulating 
	the experimental setup which has not been done in 
\cite{Comm-Evid02}. 
	The recent preprint of Feruglio et al. 
\cite{Wiss-Evid02}
	makes the same incomplete approach. 
	Calculation of the expected peak strengths starting from 
	the Table of Isotopes 
\cite{Tabl-Isot96}, 
	but {\it including} the simulation 
	gives values which are much closer to the measured intensities, 
	and in fact are consistent with them  
	within about 2 sigma experimental errors. 
	The results are given in the Table 1.\\

	From the absolute strengths we find confidence levels 
	of 3.7 and 2.6 $\sigma$ for the lines at 2010.71 
	and 2052.94 keV, respectively.
	This shows that there are lines, 
	which should not be treated as background.} \\

        {\bf That the lines are not spurious, 
	can also be seen by analyzing\cite{KK-Status02}
	the spectrum measured with natural Germanium by D. Caldwell et al. 
        more than ten years ago 
\cite{Caldw91} 
	who had the most sensitive double 
        beta experiment at that time. These authors have a three times 
        larger statistics for the background than the present experiment 
        and see essentially the same structures in the spectra, 
	as our analysis shows. 
	(As a non-enriched Ge experiment they of course 
	do not see a $0\nu\beta\beta$ signal.)\\

\newpage

\begin{table}[h]
\begin{center}
\newcommand{\m}{\hphantom{$-$}}
\renewcommand{\arraystretch}{.95}
\setlength\tabcolsep{.5pt}
\begin{tabular}{c|c|c|c|c|c|c|c}
\hline
\hline
        	          &   Intensity        &
                        &             &
        	        &      Expect.   &  Expect.   &  Aal-\\
        Energy          &       of    &
                        &       Branching       &
        Simul. of        &      rate		&	rate	
&	seth\\
                (keV)   &       Heidelberg-             &
        $\sigma$        &       Ratios\protect\cite{Tabl-Isot96}        &
        Experim.       &       accord.     &  accord.     
& et al.\protect\cite{Comm-Evid02}\\
                $*)$   &       Mos.Exper.     &
                        &       [$\%$]          &
        Setup +)       &       to sim.**) &
to\protect\cite{Tabl-Isot96}++)    &    ***)   \\
\hline
        609.312(7)      & 4399$\pm$92 &  & 44.8(5) & 
5715270$\pm$2400	&  &  &\\
        1764.494(14)    & 1301$\pm$40 &  & 15.36(20) & 
	1558717$\pm$1250	&  & &\\
        2204.21(4)      & 319$\pm$22 &  & 4.86(9) & 
	429673$\pm$656	&  & &\\
        2010.71(15)     &       37.8$\pm$10.2         &
        3.71            &       0.05(6)                 &
	15664$\pm$160	&  
	12.2$\pm$0.6	&    4.1$\pm$0.7           &       0.64    \\
        2016.7(3)       &       13.0$\pm$8.5          &
        1.53            &       0.0058(10)              &
	20027$\pm$170	&  15.6$\pm$0.7
&    0.5$\pm$0.1           &       0.08    \\
        2021.8(3)       &       16.7$\pm$8.8          &
        1.90            &       0.020(6)                &
	1606$\pm$101	&  
	1.2$\pm$0.1	&    1.6$\pm$0.5           &       0.25    \\
        2052.94(15)     &       23.2$\pm$9.0          &
        2.57            &       0.078(11)               &
	5981$\pm$115	&  
	4.7$\pm$0.3	&    6.4$\pm$1             &       0.99    \\
        2039.006        &       12.1$\pm$8.3          &
        1.46            &                               &
                        &                               &       \\
\hline
\end{tabular}
\end{center}
\caption{\label{Bi-lines}$^{214}\rm{Bi}$ is product
        of the $^{238}\rm{U}$ natural decay chain through ${\beta}^-$
        decay of $^{214}\rm{Pb}$ and $\alpha$ decay
        of $^{218}\rm{At}$.
        It decays to $^{214}\rm{Po}$ by ${\beta}^-$ decay.
        Shown in this Table are the measured intensities 
	of $^{214}\rm{Bi}$ lines
        in the spectrum shown in Fig.1 of Ref.
\protect\cite{KK-Evid01}
        in the energy window 2000 - 2060\,keV, our calculation
        of the intensities expected on the basis of the branching
        ratios given in Table of Isotopes
\protect\cite{Tabl-Isot96},
        with and without simulation of the experimental setup,
        and the intensities expected by Aalseth et al.
\protect\cite{Comm-Evid02},
	who do not simulate the setup and thus ignore summing of the
        $\gamma$ energies.
\protect\newline
        $*)$ We have considered for comparison the 3 strongest $^{214}\rm{Bi}$
        lines, leaving out the line at 1120.287\,keV (in the measured
        spectrum this line is partially overimposed on the 1115.55 keV 
        line of $^{65}\rm{Zn}$). The number of counts in each line have been
        calculated by a maximum-likelihood fit of the line with a 
        gaussian curve plus a constant background.
\protect\newline
        $+)$ The simulation is performed assuming that the impurity
        is located in the copper part of the detector chamber (best
        agreement with the intensities of the strongest lines in the
        spectrum). The error of a possible misplacement is not
        included in the calculation. The number of simulated events is 
        $10^8$ for each of our five detectors. 
\protect\newline
        $**)$ This result is obtained normalizing the simulated
        spectrum to the experimental one using the 3 strong lines
        listed in column one. Comparison to the neighboring column 
	on the right shows that the expected rates for the weak 
        lines can change strongly if we take into account the simulation.
        The reason is that the 
	line at 2010.7\,keV can be produced by summing of the 
        1401.50\,keV (1.55\%)
        and 609.31\,keV (44.8\%) lines, the one at 
        2016.7 keV by summing of the 1407.98 (2.8\%) and 609.31 (44.8\%) lines;
        the other lines at 2021.8\,keV and 2052.94\,keV 
	do suffer only very weakly from the summing effect 
	because of the different decay schemes.
\protect\newline
        $++)$ This result is obtained 
	using the number of counts for the three strong lines observed 
        in the experimental spectrum and the branching ratios from
        \protect\cite{Tabl-Isot96} without including summing effects. 
	For each
        of the strong lines the expected number of
        counts for the weak lines is calculated and then an average of 
        the 3 expectations is taken.
\protect\newline
        ***)
        Without simulation of the experimental setup.
        The numbers given here are close to those in the neighboring
        left column, when taking into account that Aalseth et al.
        refer to a spectrum which 
	contains a normalization error of a factor of 9  
	(see also point 5).
}
\end{table}


\newpage
	Thus the statement 
	in the 'Comment' about the spuriosity of the peaks has no basis.\\

	{\sf Point of Criticism 3.}
	{\it 'There is no null hypothesis analysis demonstrating 
	that the data require a peak.'}\\

	\underline{\sf Reply:}~~~ 
	{\bf This is not true. \\

	Our fit-procedure 
	allows for the case: {\it only} background, 
	line intensity {\it zero}. 
	In this sense the null hypothesis is included.}\\

	{\sf Point of Criticism 4.} 
	{\it 'There is no statement of the net counting rate of 
	the peaks other than the 2039\,keV peak.}\\

	\underline{\sf Reply:}~~~ 
	{\bf The intensities will be published in an extended paper, 
	but see Table 1.} \\

	{\sf Point of Criticism 5.} 
	{\it 'There is no presentation of the entire spectrum. 
	As a result it is difficult to compare relative peak strength.'} \\

	\underline{\sf Reply:}~~~ {\bf This is true for this Letter. 
	The full spectrum will be published in a detailed paper. 
	However, practically the same spectrum has been published 
	in a recent publication
\cite{HDM01}.
	In Fig.2 of 
\cite{HDM01}
	the spectrum is given measured with detectors 1-5 for 47.4\,kg\,y. 
	Unfortunately Fig. 1 in 
\cite{HDM01}  
	gives the spectrum measured with one detector only - {\it not} 
	of {\it all} detectors as stated there -  
	and erroneously normalized to 47.4\,kg/y. 
	The authors of 
\cite{HDM01}
	apologize for this confusing error.}\\

	{\sf Point of Criticism 6.} 
	{\it 'There are three unidentified peaks in the region 
	of analysis that have greater significance then the 2039\,keV peak. 
	There is no discussion of the origin of these peaks.'}\\

	\underline{\sf Reply:}~~~ 
	{\bf It is true that there are lines in the range beyond 
	2060\,keV, which at present cannot be identified. 
	This is, however, not relevant for the conclusions 
	concerning the signal at 2039\,keV. 
	}\\

	{\sf Point of Criticism 7.} 
	{\it 'There is no discussion of the relative peak strengths 
	before and after the single-site event cut. 
	This is needed to evaluate... the model of the peaks' origins.'}\\

	\underline{\sf Reply:}~~~ {\bf The {\it essential} point 
	can be concluded from the numbers given in the Letter
\cite{KK-Evid01}, 
	namely, that more than {\it 90\% of the signal 
	remains after the single site cut.} 
	It has been stated in 
\cite{KK-Evid01}, 
	that the analysis of the signal at 2039\,keV before correction 
	for the efficiency yields 4.6\,events (best value). 
	It has been also stated that, 
	corrected for the efficiency to identify an SSE signal,  
	the value is 8.3\,events. When normalized to the same running time as 
	in the full spectrum we obtain more than 90\% of the peak 
	contents in the full spectrum. 
	Correspondingly the half-life deduced from the single-site signal 
	at Q$_{\beta\beta}$ is (within errors) the {\it same} 
	as concluded from the full (single + multiple site signal) 
	spectra (see Table 2 in 
\cite{KK-Evid01}). 
	We add here, that in contrast the weak 
	$^{214}{Bi}$ lines are considerably reduced 
	(best values 
	to about 25\% or less. The same reduction factors are found 
	for the stronger $^{214}{Bi}$ lines e.g. at 1238, 1764, 2204\,keV 
	and, e.g., the 2614\,keV Th line.

	Note, moreover, that if the signal - consisting 
	of single site events - 
	would be due to a gamma line, it could 
        only be the double escape peak of the $\gamma$-ray, and 
	one would thus expect a 
        strong full energy peak at 2039+1022 keV. 
	Such a full energy peak is not observed 
        in the spectrum.}\\
	
	{\sf Point of Criticism 8.} 
	{\it 'No simulation has been performed to demonstrate 
	that the analysis correctly finds true peaks or that 
	it would find no peaks if none existed. 
	Monte Carlo simulations of spectra with varying numbers 
	of peaks confirming the significance of found peaks are needed.'}\\

	\underline{\sf Reply:}~~~ {\bf This it not true}.\\
	
	{\bf Of course, such simulations have been performed 
\cite{DARK2002,DM2002}.
	They are important to prove the correctness of our computer programs   
	- but not in the sense that one would have to prove the  
	Bayesian or Maximum Likelihood methods, 
	which are well established. 

	As mentioned in the reply to Question 1, we have made numerical 
	simulations  
	which e.g. show, on the basis of 1000 simulated spectra 
	containing {\it no} line, that the probability to find a line 
	originating from statistical fluctuations,
	at a given energy {\it above} a confidence level of 95$\%$,  
	is about 4.2 percent.

	The simulations thus show, that our analysis programs calculate
	the probabilities for the existence of a line 
	in a correct way. 
	In particular they confirm that the signal at Q$_{\beta\beta}$  
	can be faked by statistical fluctuations only 
	with the small probability (1 - K$_E$), $\sim$ 3$\%$.

	An important point of the analysis is, that we can use the 
	{\it exact energy position} of the line 
	(Q$_{\beta\beta}$ is known with very high accuracy 
	to be 2039.006(50)\,keV) {\it and} the {\it width of} 
	the line (determined from known strong  
	$\gamma$-lines in the spectrum), 
	as well as its {\it shape} (Gaussian), 
	as an {\it input} into 
	the search procedure. This is the reason, that the method 
	can do {\it more the naked eye}.}\\

	{\sf Point of Criticism 9.} 
	{\it 'There is no discussion of how sensitive the conclusions 
	are to different mathematical models. There is a previous 
	Heidelberg-Moscow publication that gives a lower 
	limit of 1.9$\times{10}^{25}$ y (90$\%$ confidence level). 
	This is in conflict with the ``best value'' 
	of the new KDHK paper 1.5$\times {10}^{25}$ y. 
	This indicates a dependence of the results on the analysis 
	model and the background evaluation'.}\\

	\underline{\sf Reply:}~~~  {\bf This is not true.} \\

	{\bf There is {\it no} discrepancy between 
	the results obtained in\cite{KK-Evid01} and\cite{HDM01}.  
	In\cite{HDM01}
	we exclude on a 90$\%$ c.l., with the method used there, 
	the number of counts

	N $<$ 19.8\hspace{2.cm} (full spectrum)

	N $<$  ~9.3\hspace{2.cm} (SSE spectrum)

\noindent
	With the data in Mod. Phys. Lett. A, with the Bayesian analysis, 
	we get at 90$\%$ c.l., when using the {\it same} energy 
	window of analysis (2000 - 2080\,keV) and assuming 
	all structures in this range to be background

	N $<$ 15.0\hspace{2.cm} (full spectrum)

	N $<$  ~8.3\hspace{2.cm} (SSE spectrum)

	According to the Particle Data Group 
	(Eur. Phys. J. C15 (2000) 1)  
	some slight dependence of the result on the method 
	of analysis would be not surprising.}}\\

	Summarizing, the criticism made in the 'Comment'
\cite{Comm-Evid02}
	is, in view of the Replies given here, 
	not justified in any of the points raised 
	(the same is true, 
	from the same arguments, for the criticism raised recently in 
\cite{Harn-Id02}).\\

	We think that it remains useful and inspiring to have informed 
	the neutrino community about our evidence for 
	a 2.2$\sigma$ - 3.1$\sigma$ result 
	on the 0$\nu\beta\beta$ decay.\\

\noindent
	{\large\bf Acknowledgment}
\vspace{0.5cm}

	The author thanks A. Dietz, I.V. Krivosheina and C. Tomei 
	for useful discussions.


\end{document}